\documentclass[
 aps,
 prd,
twocolumn,
superscriptaddress,
frontmatterverbose, 
showpacs,preprintnumbers,
 nofootinbib,
 amsmath,
 amssymb,
floatfix,
latexsym,array,enumerate,letter,
]{revtex4-1}

\usepackage[utf8]{inputenc}
\usepackage[english]{babel}
\usepackage{multirow}
\allowdisplaybreaks
\usepackage{graphicx,color}
\usepackage[colorlinks=true,citecolor=blue,linkcolor=blue,urlcolor=blue]{hyperref}
\usepackage{comment}
\usepackage{appendix}
\usepackage{caption}
\usepackage[normalem]{ulem}

\usepackage{subcaption}
\usepackage{footmisc}
\usepackage{bm}
\usepackage{dcolumn}
\usepackage{gensymb}
 \usepackage{float} 
\restylefloat{table}
\usepackage{multirow}
 \usepackage{array}
 
\usepackage{subcaption}

\hypersetup{
    colorlinks,
    citecolor=blue,
    filecolor=black,
    linkcolor=blue,
    urlcolor=blue
}

\newcommand{\be}{\begin{equation}}
\newcommand{\ee}{\end{equation}}
\newcommand{\beq}{\begin{equation}}
\newcommand{\eeq}{\end{equation}}
\newcommand{\bea}{\begin{eqnarray}}
\newcommand{\eea}{\end{eqnarray}}

\usepackage{graphicx}
\usepackage{caption}
\usepackage{subcaption}

\usepackage{xcolor}

\begin{document}

\title{Hybrid Quark Stars with Quark-Quark Phase Transitions}

\author{Zongcan Yang}
\affiliation{Department of Astronomy, School of Physics, Huazhong University of Science and Technology, Wuhan, 430074, China}

\author{Tianxiong Zeng}
\affiliation{Department of Astronomy, School of Physics, Huazhong University of Science and Technology, Wuhan, 430074, China}

\author{Yan Yan}
\affiliation{Wang Zheng School of Microelectronics, Changzhou University, Changzhou 213164, China}

\author{Wen-Li Yuan}
\affiliation{School of Physics and State Key Laboratory of Nuclear Physics and Technology, Peking University, Beijing 100871, China}

\author{Chen Zhang}
\email{zhangvchen@tongji.edu.cn}
\affiliation{School of Physics Science and Engineering, Tongji University, Shanghai 200092, China}
\affiliation{The HKUST Jockey Club Institute for Advanced Study,
The Hong Kong University of Science and Technology, Hong Kong, P.R. China}

\author{Enping Zhou}
\email{ezhou@hust.edu.cn}
\affiliation{Department of Astronomy, School of Physics, Huazhong University of Science and Technology, Wuhan, 430074, China}

\begin{abstract}
We explore the possibility of phase transitions between different quark matter phases occurring within quark stars, giving rise to the hybrid quark stars (HybQSs). First, we obtain the generic phase diagram of possible mass-radius relation forms for the HybQSs. Then, utilizing a well-established general parameterization of interacting quark matter, we construct quark star models featuring sharp first-order quark-quark phase transitions of various types, in contrast to the hadron-quark transition in conventional hybrid stars. 
We systematically investigate how recent observations, such as the pulsar mass measurement $M_{\rm TOV}\gtrsim2M_{\odot}$ and the GW170817's tidal deformability bound $\Lambda_{1.4M_{\odot}}<800$, constrain the viable parameter space. We also identified twin stars in some of the HybQS parameter space. Furthermore, we found that the quark-quark phase transitions in hybrid quark stars may also cause the deviation from the approximate universal relation between the dominant postmerger frequency $f_{\rm peak}$ and tidal deformability $\Lambda_{1.35M_{\odot}}$, which was previously believed to only be caused by the hadron-quark phase transitions in hybrid neutron stars. This work unveils new possibilities of phase transitions and the resulting new types of compact stars in realistic astrophysical scenarios.

\end{abstract}

\maketitle

\section{Introduction}
\label{sec:level1}

Recent detections of gravitational waves (GWs) from the coalescence of compact binaries by LIGO/Virgo collaborations~\cite{LIGOScientific:2016aoc, LIGOScientific:2017bnn, LIGOScientific:2018mvr, LIGOScientific:2017vwq, LIGOScientific:2018hze,LIGOScientific:2020aai, LIGOScientific:2020zkf, LIGOScientific:2018cki} have greatly advanced our knowledge and probing ability on black holes and compact stars, the density regime of which involves strong interaction physics that suffers from its non-perturbative nature. The potential occurrence of phase transitions within compact stars represents a long-standing subject of nuclear astrophysics \cite{doi:10.1142/S0218301318300084,orsaria2019phase,heiselberg1999phase,glendenning2001phase,yang2008influence,reddy2000first,glendenning1992first,burgio2002hadron,Weih:2019xvw,annala2020evidence}. Numerous studies have explored the possible formation of deconfined quark matter (QM) in the core region of neutron stars that potentially arises via a hadron-quark phase transition. Such a transition would lead to the formation of hybrid stars~\cite{Alford:2004pf,Alford:2013aca}, which has some distinct astrophysical signatures~\cite{Alford:2019oge,bauswein2019identifying} and may help explain various pulsar observations~\cite{Paschalidis:2017qmb,Nandi:2017rhy,Montana:2018bkb,Blaschke:2022egm,Contrera:2022tqh,Yuan:2025dft,Yuan:2024nho,Gao:2024lzu,Wang:2019npj,Christian:2018jyd,Zhao:2022tcw,Constantinou:2023ged,Ghosh:2024auj,Ghosh:2024auj,Alford:2013aca,annala2020evidence,Li:2024rua}.

An alternative scenario emerges if quark matter exhibits absolute stability even at zero pressure. This could manifest as three-flavor strange quark matter (SQM)~\cite{Bodmer:1971we,witten1984cosmic,Terazawa:1979hq,Farhi:1984qu} or two-flavor up-down quark matter ($ud$QM)~\cite{Holdom:2017gdc}. In such cases, quark matter could constitute the entirety of the compact star, forming strange quark stars or up-down quark stars~\cite{Haensel:1986qb,Alcock:1986hz,olinto1987conversion,Xu:1999bw,Zhou:2017pha,Zhang:2019mqb,Ren:2020tll,Cao:2020zxi,Wang:2021byk,Yuan:2022dxb,Restrepo:2022wqn,Xia:2020byy,Xia:2022tvx,Yang:2023haz,Su:2024znh,Zhou:2024syq,Xie:2025sth}. In this context, it is natural to explore phase transitions within such quark stars. For example, as density increases, it's commonly expected that unpaired quark matter may form color-superconducting Cooper pairs due to the attractive channel in the one-gluon exchange~\cite{Alford:2007xm,Alford:2017qgh}, either in two-flavor color superconductivity, where $u$ quarks pair with $d$ quarks [conventionally termed ``2SC" (``2SC+s") without (with) unpaired strange quarks], or in a color-flavor locking (CFL) phase, where $u,d,s$ quarks pair with each other antisymmetrically.  Besides,  the feedback of quark gas on the quantum chromodynamics (QCD) vacuum tends to turn on the strangeness to lower the energy budget when density increases beyond some large threshold value~\cite{Holdom:2017gdc}, inducing a two-flavor QM to three-flavor QM phase transition.
Therefore, it is natural to raise the new possibility of hybrid quark stars (HybQS), which are composed entirely of quark matter but in different phases in core and crust, arising from the phase transition between the two different quark matter phases (QM-QM transition). 

To our knowledge, no precedent study has explored this quark-quark phase transition in the absolutely stable quark matter context and the resulting HybQS. References ~\cite{Alford:2017qgh,Li:2023zty,Sedrakian:2023eqv} studied sequential phase transitions that involve quark-quark phase transitions but as the secondary phase transition, so that their object is the conventional hybrid star, which has a hadronic matter crust, while HybQS is self-bounded by its quark matter crust. Other self-bound hybrid stars include hybrid strangeon stars~\cite{Zhang:2023szb}  from the quark cluster matter~\cite{Xu:2003xe,Lai:2017ney,Zhang:2023mzb} to quark matter phase transition, and inverted hybrid stars~\cite{Zhang:2022pse,Zhang:2023zth,Negreiros:2024cvr} resulting from a quark matter to hadronic matter transition. These are all very different from HybQSs where only different phases of quark matter are involved.

Pure quark stars are known to form either when neutron stars absorb quark matter nuggets or via quantum nucleation in the interior~\cite{olinto1987conversion,Bombaci:2016xuj,Ren:2020tll}.  If one phase of quark matter becomes more stable than another above some approachable density, quantum nucleation can subsequently occur within these quark stars, creating hybrid quark stars. This phase transition requires the central pressure to exceed a critical value at the corresponding central chemical potential. Such an increase in central pressure beyond the critical point can result from spin-down, accretion, or merger of the quark stars.

As for the organization of this paper, first, we introduce the equation of state (EOS)  models adopted in constructing HybQSs, and discuss possible phase transitions. Then, we explore the phase diagrams for the mass-radius relation of HybQSs. Next, we examine the viable parameter space of HybQSs in different phase transition types against astrophysical tests, including the requirement on maximum mass $M_{\rm TOV} \gtrsim 2 M_{\odot}$~\cite{demorest2010shapiro}, and the tidal deformability constraint $\Lambda_{1.4M_{\odot}}<800$ from LIGO/Virgo GW170817 event~\cite{LIGOScientific:2017vwq}. After that, we demonstrate that quark–quark phase transitions can also result in deviation from the approximate universal relation between $f_{\text{peak}}$ and $\Lambda_{1.35M_{\odot}}$.

\section{Quark Matter EOS Model}
\label{sec:level4}
We employ the widely used interacting quark matter EOS \cite{zhang2021unified}, which has the simple form
\begin{equation}
P=\frac{1}{3}(\rho-4B)+\frac{4\lambda^2}{9\pi^2}\left(-1+\mathrm{sgn}(\lambda)\sqrt{1+3\pi^2\frac{(\rho-B)}{\lambda^2}}\right),
\label{EOS}
\end{equation}
where the gap parameter $\Delta$ for the color-superconducting quark phase, the strange quark mass $m_s$, and the parameter $a_4$ representing perturbative QCD corrections are encapsulated into a single parameter $\lambda$:
\be
\lambda=\frac{\xi_{2a} \Delta^2-\xi_{2b} m_s^2}{\sqrt{\xi_4 a_4}},
\label{lam}
\ee which characterizes the strength of strong interaction effects. $B$ is the effective bag constant that depicts the QCD vacuum contributions. Considering the feedback of quark gas on the QCD vacuum and the fact that $\rm SU(3)$$_f$ flavor symmetry is broken by the strange quark mass, Ref.~\cite{Holdom:2017gdc} has shown that two-flavor quark matter has a smaller $B$ than the three-flavor case, which we adopt as a requirement in the following study. Since color-superconductivity mainly affects the quark gas rather than the QCD vacuum, we assume QM with the same flavor composition shares the same bag constant value.

The parameters for different quark matter phases in Eq. (\ref{EOS}) are:
\begin{equation}
(\xi_4, \xi_{2a}, \xi_{2b}) =
\begin{cases}
\left(\left( \left( \frac{1}{3} \right)^{\frac{4}{3}} + \left( \frac{2}{3} \right)^{\frac{4}{3}} \right)^{-3}, 0, 0\right) & \text{unpaired 2$f$}\\
\left( \left(\left( \frac{1}{3} \right)^{\frac{4}{3}} + \left( \frac{2}{3} \right)^{\frac{4}{3}} \right)^{-3}, 1, 0  \right) & \text{2SC} \\
(3, 0, 3/4) & \text{unpaired 3$f$}
\\
(3, 1, 3/4) & \text{2SC+s} 
 \\
(3, 3, 3/4) & \text{CFL}
\end{cases}
\label{types}
\end{equation}
where unpaired 2$f$ and 3$f$ means two-flavor and three-flavor quark matter without color superconductivity. 
 The chemical potential for quark matter is \cite{zhang2021unified}:
\begin{equation}
\mu_{\mathrm{Q}}=\frac{3\sqrt{2}}{(a_4\xi_4)^{1/4}}\sqrt{[(P+B)\pi^2+\lambda^2]^{1/2}-\lambda}\ .
\label{chem}
\end{equation}
Taking zero pressure limit, we obtain the energy per baryon number
\be
E/A=\frac{3\sqrt{2}}{(a_4\xi_4)^{1/4}}\sqrt{[B\pi^2+\lambda^2]^{1/2}-\lambda}\ .
\label{EA}
\ee
Note that $a_4$ and $\xi_4$ parameters, although do not enter the EOS in Eq. (\ref{EOS}), affect the chemical potential, and thus the phase transition pressures, which are determined by the crossing of the chemical potentials of two quark matter phases. In this proof-of-concept work, we neglect perturbative QCD corrections ($a_4=1$). Besides, we adopt the dimensionless normalization form  
\be
\bar{\lambda}=\lambda^2/(4B)\ ,
\label{lambar}
\ee
so that the EOS in Eq. (\ref{EOS}) is determined by two parameters $(\bar{\lambda}, B)$.  We consider Maxwell constructions for various sharp quark-quark phase transitions (no mixed phases). 
The net EOS in our hybrid quark star model includes 4 independent parameters: the crust quark matter phase parameters $B_{\rm crust}$, $\bar{\lambda}_{\rm crust}$ and the core quark matter phase parameters $B_{\rm core}$, $\bar{\lambda}_{\rm core}$, with the transition pressure $P_{\text{trans}}$ determined by the crossing of the chemical potentials $\mu$ of the two QM phases. The net EOS is then:
\begin{equation}
\rho(P)=
\begin{cases}
\rho(P,\bar{\lambda}_{\rm crust},B_{\rm crust})\ , & P<P_{\mathrm{trans}} \\
\rho(P,\bar{\lambda}_{\rm core},B_{\rm core})\ ,&P>P_{\mathrm{trans}}
\end{cases}
\end{equation}
With the EOS above, we solve the Tolman-Oppenheimer-Volkov (TOV) equations \cite{tolman1939static,oppenheimer1939massive}:
\begin{equation}
\begin{aligned}
\frac{dP(r)}{dr} &= -\frac{\left[ m(r) + 4\pi r^3 P(r) \right] \left[ \rho(r) + P(r) \right]}{r \left( r - 2m(r) \right)}\ , \\
\frac{dm(r)}{dr} &= 4\pi \rho(r) r^2\ ,
\label{tov}
\end{aligned}
\end{equation}
where $P$ is pressure and $m$ is physical mass. Using the central pressure $P_{\text{center}}$ as the initial condition and the boundary condition $P(R)=0$, we obtain the radius $R$ and stellar mass $M=m(R)$, yielding the mass-radius relation for hybrid quark stars.

The dimensionless tidal deformability $\Lambda=2k_2/(3C^5)$, where $C=M/R$ is the compactness and $k_2$ is the tidal-response Love number~\cite{hinderer2008tidal,hinderer2010tidal,postnikov2010tidal}, is computed for comparison with gravitational wave observations. Determining $k_2$ requires solving a differential equation for $y(r)$~\cite{postnikov2010tidal} concurrently with the TOV equation (Eq.~(\ref{tov})), using $y(0)=2$ as the boundary condition. In the case of self-bound hybrid stars, an additional matching condition $y(r_{d}^+) - y(r_{d}^-) = -4\pi r_{d}^3 \Delta \rho_d /(m(r_{d})+4\pi r_{d}^3 P(r_{d}))$ is necessary at locations $r_d$ of energy density jumps $\Delta \rho_d$~\cite{damour2009relativistic,takatsy2020comment}; these locations include both the core radius $r_{\rm core}$ and the whole star radius $R$.

\section{$M$-$R$ Phase Diagram of HybQSs}
Twin stars with the same mass but different radii provide additional possibilities and strong constraints on the underlying EOS. The possible existence of twin stars has attracted considerable attention in the research of phase transition in compact stars. Therefore, we analyze the phase diagrams for the mass-radius relation of HybQSs. We first use the simplest HybQSs model: the MIT+CSS model. This model has the outer quark phase described by the MIT bag model, while the inner quark phase is described by the constant sound speed (CSS) model. In general, we can obtain the corresponding EOS:
\begin{equation}
\rho(P)=
\begin{cases}
3P+4B\ , & P<P_{\mathrm{trans}} \\
\rho_{\rm trans}+\Delta\rho+c_s^{-2}(P-P_{\rm trans})\ ,&P>P_{\mathrm{trans}}
\end{cases}
\end{equation}
where $c_{\rm s}$ is the sound speed. The whole model is described by four independent parameters ($B$, $P_{\rm trans}$, $\Delta\rho$, $c_{\rm s}^{2}$). Considering that the outer layer quark matter needs to satisfy $E/A<930$ MeV to meet the absolute stability against bulk nuclear matter, and the fact that the MIT bag EOS can be obtained by taking zero $\lambda$ limit of the interacting quark matter EOS (Eq.~(\ref{EOS})), we can obtain a necessary upper limit of the outer-layer bag constant from $E/A<930$ MeV by substituting zero $\lambda$ and unity $a_4$ limit of Eq. (\ref{EA}). Thus,
 $B<(\frac{\xi_4^{1/4}}{3\sqrt{2\pi}}930 \text{ MeV})^4$, yielding $B\lesssim(162.8 \text{ MeV})^4\approx91\, \rm MeV/fm^3$ for absolutely-stable SQM \footnote{Note that in the conventional absolutely-stable SQM picture, the $ud$QM is unstable, which translates to a lower bound $B\gtrsim (144.38 \text{ MeV})^4\approx 56.8\, \rm MeV/fm^3$, but this lower bound can be relaxed significantly by a smaller $a_4$ as indicated by Eq.~(\ref{EA}) without affecting the EOS (since $a_4$ does not enter the MIT bag EOS expression derived from the $\bar{\lambda}\to 0$ limit of Eq. (\ref{EOS}).}, and $B\lesssim (144.38 \text{ MeV})^4\approx 56.8\, \rm MeV/fm^3$ for absolutely-stable $ud$QM.

\begin{figure*}[htb]
\captionsetup[subfigure]{justification=centering}
    \centering
    \begin{subfigure}[b]{0.45\linewidth}
    \includegraphics[width=\linewidth]{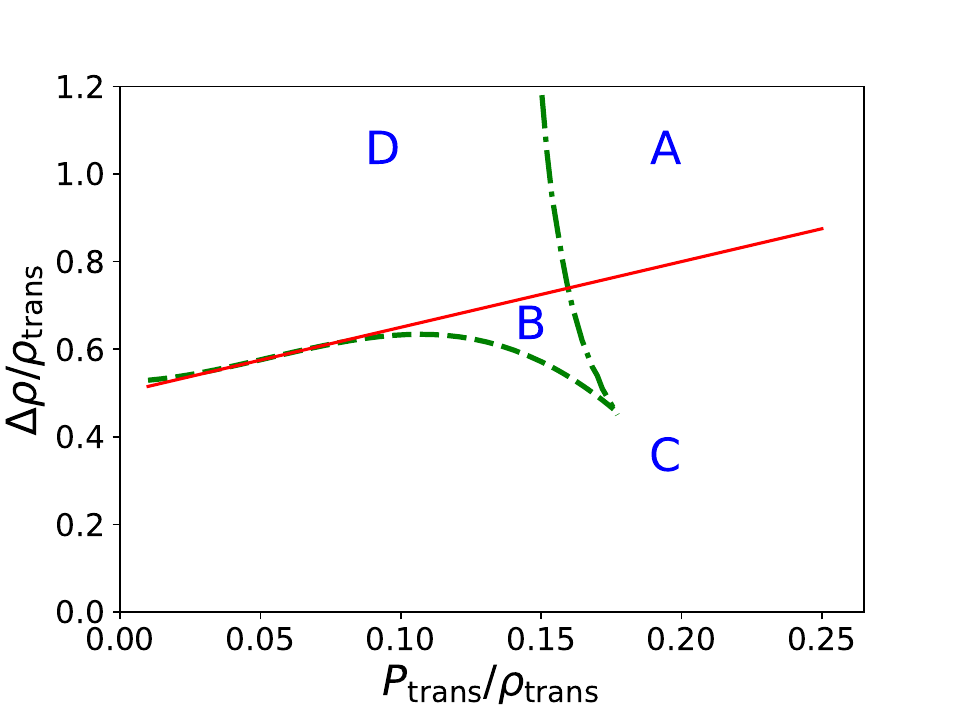}
    \caption{}
    \label{fig:a}
    \end{subfigure}
    \begin{subfigure}[b]{0.45\linewidth}
    \includegraphics[width=\linewidth]{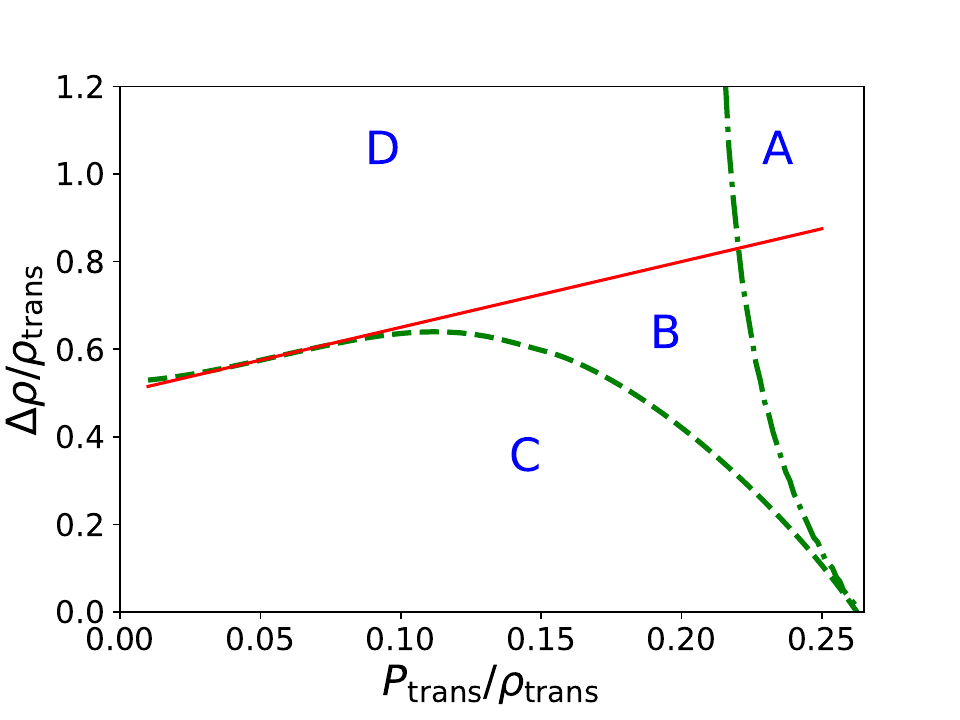}
    \caption{}
    \label{fig:b}
    \end{subfigure}
    \caption{Phase diagram for the mass-radius relation of HybQS branches. The red line corresponds to Eq. (\ref{critical_rho}). In (a), we fix the parameters as $B$ = 50 {MeV}/{fm}$^3$  and  $c_{s}^{2}$ = 1/3. In (b), we fix the parameters as $B$ = 50 MeV/{fm}$^3$  and  $c_{s}^{2}$ = 1. The mass–radius relations for regions A, B, C, and D are presented in Figs. \ref{MITCSS_MR} (a), (b), (c) and (d), respectively.}
    \label{MITCSS_phasediagram}
\end{figure*}

\begin{figure*}[htb]
\captionsetup[subfigure]{justification=centering}
    \centering
    
    \begin{subfigure}[b]{0.45\linewidth}
        \includegraphics[width=\linewidth]{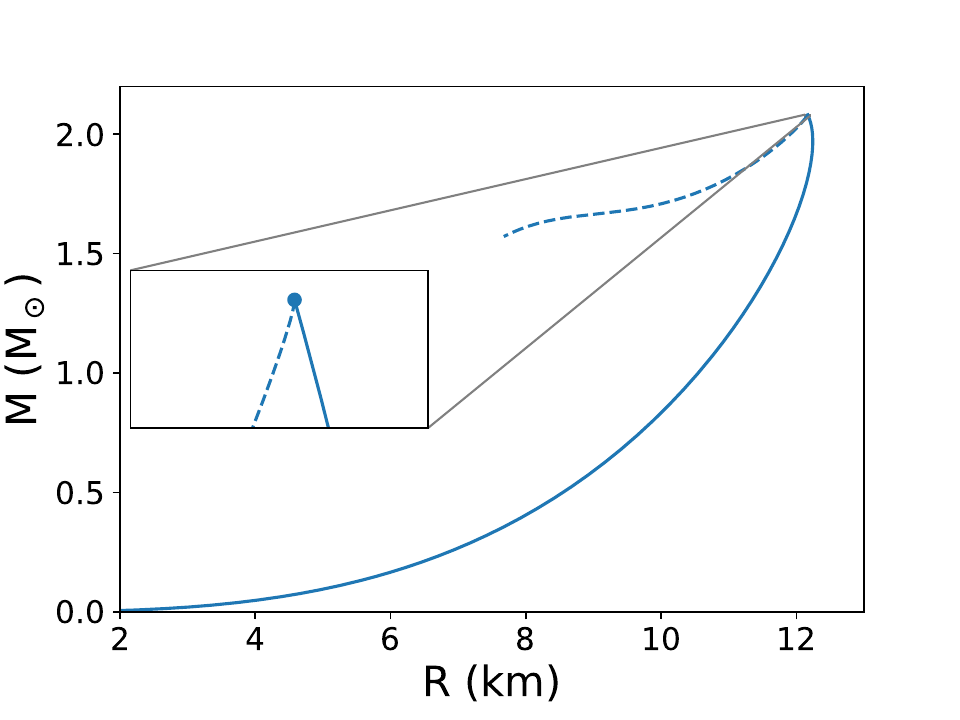}
        \caption{}
        \label{fig:a}
    \end{subfigure}
    \hfill
    \begin{subfigure}[b]{0.45\linewidth}
        \includegraphics[width=\linewidth]{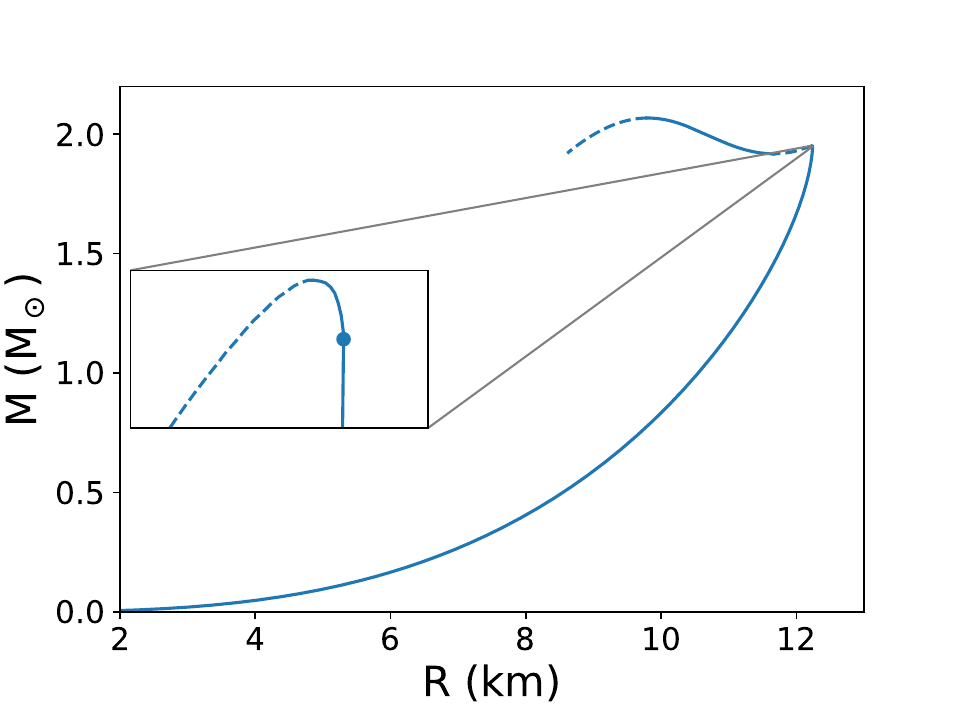}
        \caption{}
        \label{fig:b}
    \end{subfigure}
    
    \begin{subfigure}[b]{0.45\linewidth}
        \includegraphics[width=\linewidth]{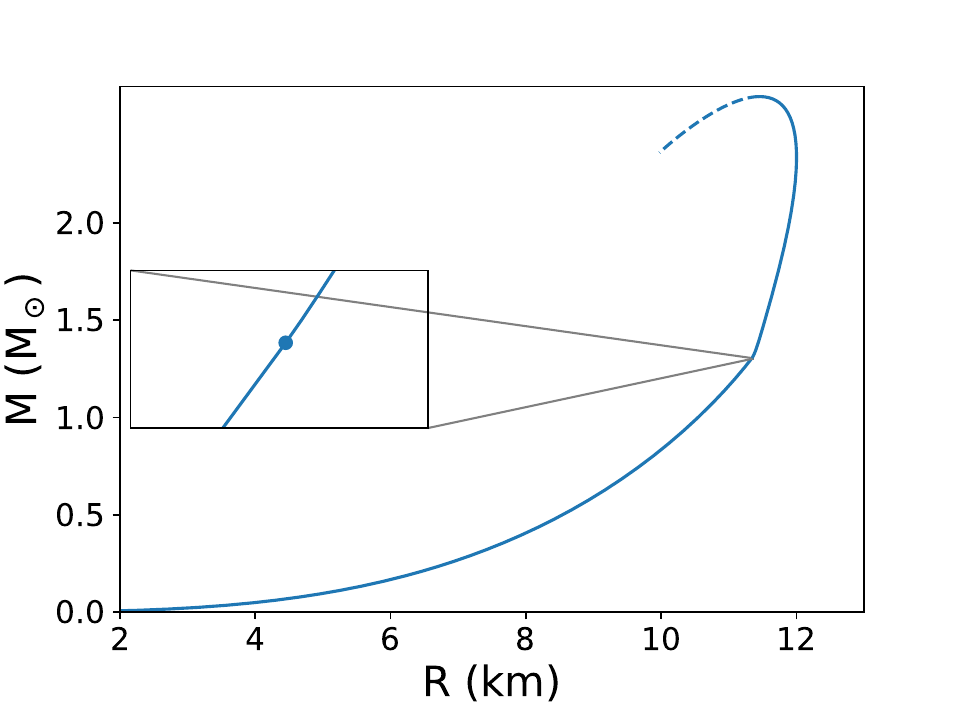}
        \caption{}
        \label{fig:c}
    \end{subfigure}
    \hfill
    \begin{subfigure}[b]{0.45\linewidth}
        \includegraphics[width=\linewidth]{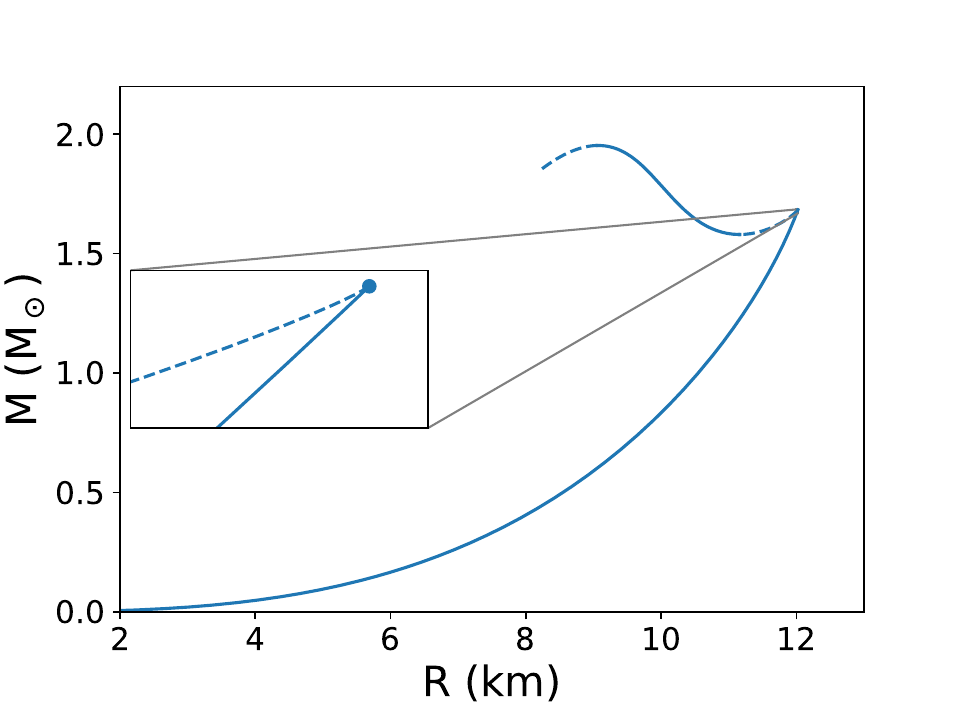}
        \caption{}
        \label{fig:d}
    \end{subfigure}
    
    \caption{$M$-$R$ relations of typical HybQS examples in different regions of Fig. \ref{MITCSS_phasediagram}. The insets on the left present enlarged views of the vicinity of the phase transition points, with the phase transition points indicated by blue dots. (a), (b), (c) and (d) correspond to regions A, B, C and D in Fig. \ref{MITCSS_phasediagram}, respectively.
}
    \label{MITCSS_MR}
\end{figure*}

For the hadron-quark hybrid stars, a small energy density jump $\Delta\rho$ leads to a stable hybrid star branch connected to the hadronic star branch. The critical value $\Delta\rho_{\mathrm{crit}}$ is given by ~\cite{seidov1971stability,Alford:2013aca}:
\begin{equation}
\frac{\Delta\rho_{\mathrm{crit}}}{\rho_{\mathrm{trans}}}=\frac{1}{2}+\frac{3}{2}\frac{p_{\mathrm{trans}}}{\rho_{\mathrm{trans}}}\ .
\label{critical_rho}
\end{equation}
We explore the $M$-$R$ phase diagrams of HybQSs, as shown in Fig.~\ref{MITCSS_phasediagram}, and find that Eq.~(\ref{critical_rho}) is also applicable. There are four types of mass-radius relations in Fig.~\ref{MITCSS_phasediagram}. In region A, there is no stable hybrid quark star branch. In region B, the mass–radius relation exhibits both connected and disconnected branches. In region C, only a connected branch exists. In region D, there always exists a disconnected branch. In regions B and D, there are twin star solutions. These general features are explicitly demonstrated in Fig.~\ref{MITCSS_MR}. These results have very similar qualitative behavior compared to the hadron-quark hybrid stars results regarding the mass-radius evolution after the transition, as discussed in Ref. \cite{alford2013generic}.  

Although the MIT+CSS model provides a general parameter scale, it lacks a direct connection to underlying physics. In the following, we adopt a more physically motivated model as introduced in the previous section to investigate the properties of HybQSs.

\section{HybQS Construction}

Both the number of quark flavors and pairing properties could be different in the crust and core phases inside a HybQS. We explore all possible HybQS models, as listed in the following 3 categories according to the flavors in each phase:
\begin{itemize}
\item  $2f \rightarrow 2f$: unpaired 2$f$ to 2SC
\item  $3f \rightarrow 3f$: unpaired 3$f$ to 2SC+s or CFL; 2SC+s to CFL 
\item  $2f \rightarrow 3f$: unpaired 2$f$ to unpaired 3$f$ or 2SC+s or CFL;  2SC to 2SC+s or CFL
\end{itemize}

We examine the viable EOS parameter space under theoretical considerations described as follows:
\begin{enumerate}
\item The energy per baryon number ($E/A$) for the quark matter at the crust layer should be absolutely stable as the true ground state, i.e., $E/A=\mu_{\rm crust}(P=0) < 930$ MeV. 
\item The crust phase is also more stable than the core phase at low pressure, i.e., $\mu_{\rm crust}(P=0) < \mu_{\rm core}(P=0)$.
\item We only consider cases where the chemical potential curves intersect once as pressure increases.
\item The effective bag constant satisfies $B_{\rm crust} < B_{\rm core}$ for $2f \rightarrow 3f$ cases, while $B_{\rm crust} = B_{\rm core}$ for $2f \rightarrow 2f$ and $3f \rightarrow 3f$ cases, as previously reasoned.
\item  The central pressure of the maximum-mass star must meet $P_{\rm TOV}>P_{\rm trans}$, to avoid the scenario in which all stellar configurations with a phase transition are unstable.

\end{enumerate}
Satisfying these conditions yields viable EOSs for hybrid quark stars.

We then constrain the parameter space using two widely used astrophysical constraints: 
\begin{enumerate}
    \item Maximum mass $M_{\text{TOV}} \gtrsim 2M_\odot$~\cite{demorest2010shapiro};
    \item Tidal deformability $\Lambda_{1.4M_{\odot}} < 800$ inferred from the GW170817 event~\cite{LIGOScientific:2017vwq}.
\end{enumerate}

\subsection{$2f \rightarrow 2f$ and $3f \rightarrow 3f$}
For the case that flavors of quarks stay the same in the crust and core phases, we find no parameter space for HybQS solutions if the bag constant is only flavor-sensitive, which is a natural assumption considering the physical meaning of the model parameters. To show the explicit reason, we can take the derivative of Eq.~(\ref{chem}) and obtain
\begin{equation}
\frac{d\mu}{d\lambda}=-\frac{3\sqrt{2}}{2\xi_4^{1/4}}\cdot\frac{\left(\sqrt{(P+B)\pi^2+\lambda^2}-\lambda\right)^{1/2}}{\sqrt{(P+B)\pi^2+\lambda^2}}
\label{Chemical potential derivative}
\end{equation}
which shows $d\mu/d\lambda < 0$ always holds, so that for shared $B$ and $\xi_4$ for the same-flavor QM phases, $\mu_{\rm crust}(P) < \mu_{\rm core}(P)$ always holds if $\mu_{\rm crust}(0) < \mu_{\rm core}(0)$, preventing an intersection. 

\subsection{Unpaired $2f \rightarrow$ Unpaired $3f$}
\begin{figure*}[htb]
\captionsetup[subfigure]{justification=centering}
    \centering
    \begin{subfigure}[b]{0.45\linewidth}
    \includegraphics[width=\linewidth]{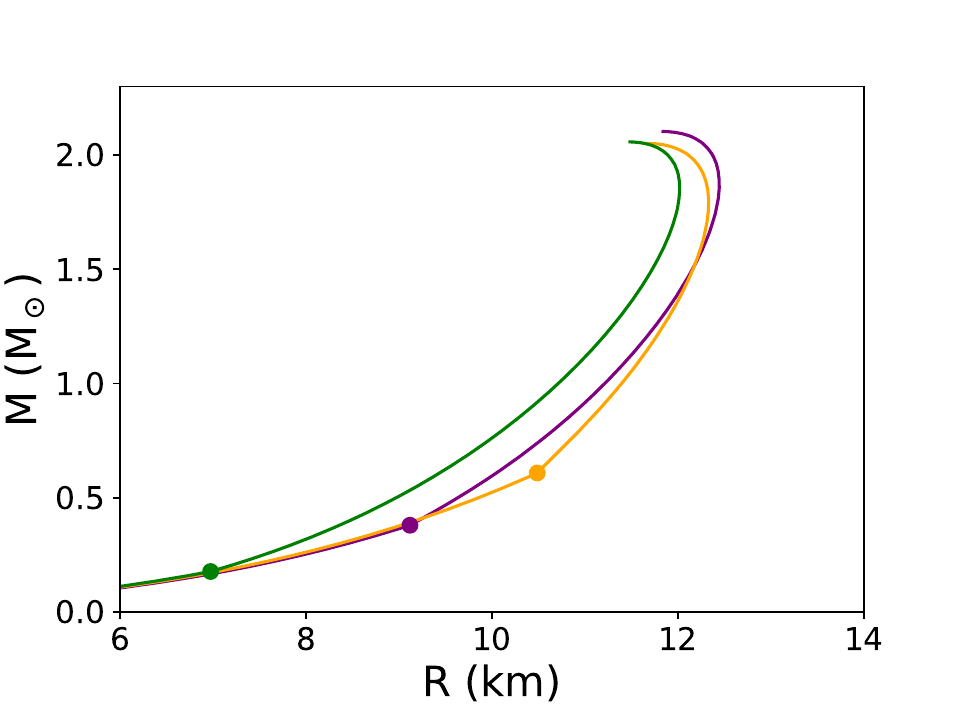}
    \caption{}
    \label{fig:a}
    \end{subfigure}
    \begin{subfigure}[b]{0.45\linewidth}
    \includegraphics[width=\linewidth]{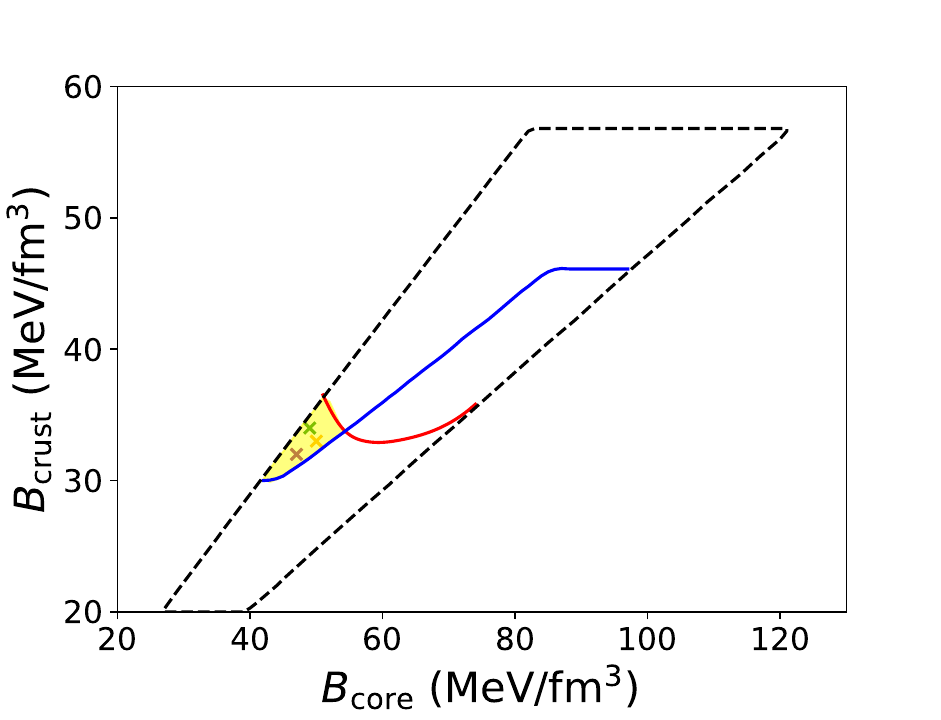}
    \caption{}
    \label{fig:b}
    \end{subfigure}
    \caption{(a) $M$-$R$ relations of HybQS benchmark examples with unpaired 2$f$ to unpaired 3$f$ phase transition. (b) related parameter space, where the cross markers correspond to the same-color $M$-$R$ in (a). In (b), the gray dashed contour encloses the parameter space consistent with theoretical considerations. The red line corresponds to the $M_{\rm TOV} = 2\,M_\odot$ constraint, below which one has $M_{\rm TOV} \gtrsim 2\,M_\odot$, while the blue line represents the $\Lambda_{1.4M_{\odot}} = 800$ constraint, above which one has $\Lambda_{1.4M_{\odot}} < 800$. The enclosed yellow region indicates the intersection of these two observational constraints, highlighting the parameter space simultaneously satisfying both.
}
    \label{udQMtoSQM}
\end{figure*}

Next, let's consider HybQSs with the transition from unpaired 2$f$ to unpaired 3$f$. Referring to Eq.~(\ref{lam}) and Eq.~(\ref{lambar}), for unpaired 2$f$ we have $\lambda_{\rm crust} = \bar{\lambda}_{\rm crust} = 0$, while for unpaired 3$f$ with $m_s = 100~\mathrm{MeV}$ we obtain $\lambda_{\rm crust} \approx -4330~\mathrm{MeV}^2$. Thus, the model effectively reduces from four parameters to two parameters $(B_{\mathrm{core}},B_{\mathrm{crust}})$. In Fig.~\ref{udQMtoSQM}a, we show the mass-radius relations of benchmark examples for illustration of the typical feature. 
Besides, as the top black-dashed line of Fig.~\ref{udQMtoSQM}b shows, HybQS solutions of this type should have $B_{\mathrm{crust}} \lesssim 56.8 \rm\,MeV/fm{^3}$ as determined from the absolute stability condition $\mu(P=0) < 930$ MeV, being consistent with the result of pure quark stars composed of unpaired $2f$ (Eq. (2) of Ref.~\cite{Zhang:2019mqb}). The rightmost black-dashed boundary arises from the fact that no stable HybQS branch exists beyond this line. The left boundary is set by relative stability at zero pressure ($\mu_{\rm crust}(P=0) < \mu_{\rm core}(P=0)$).  For the lines of astrophysical constraints, we can see that in the $\Lambda_{1.4M_{\odot}} = 800$ line, there is a horizontal segment at $B_{\mathrm{crust}} \approx 46\rm\, {MeV/fm^3}$. This is because when $B_{\mathrm{core}} > 86\rm \, {MeV/fm^3}$, the phase transition always occurs at a mass larger than $1.4\,M_\odot$. Therefore $\Lambda_{1.4M_{\odot}}$ depends only on $B_{\mathrm{crust}}$, and hence independent of $B_{\mathrm{core}}$. The critical value $B_{\mathrm{crust}} \approx 46\rm\, {MeV/fm^3}$ determined by $\Lambda_{1.4M_{\odot}}=800$ is consistent with the projection of $\Lambda_{1.4M_{\odot}}=800$ line on the bag constant axis at zero $\bar{\lambda}$ in Fig. 3 of Ref.~\cite{zhang2021unified}. 

\subsection{Unpaired $2f \rightarrow$ 2SC+s and Unpaired $2f \rightarrow$ CFL}
\begin{figure*}[htb]
\captionsetup[subfigure]{justification=centering}
    \centering
    \begin{subfigure}[b]{0.4\linewidth}
    \includegraphics[width=\linewidth]{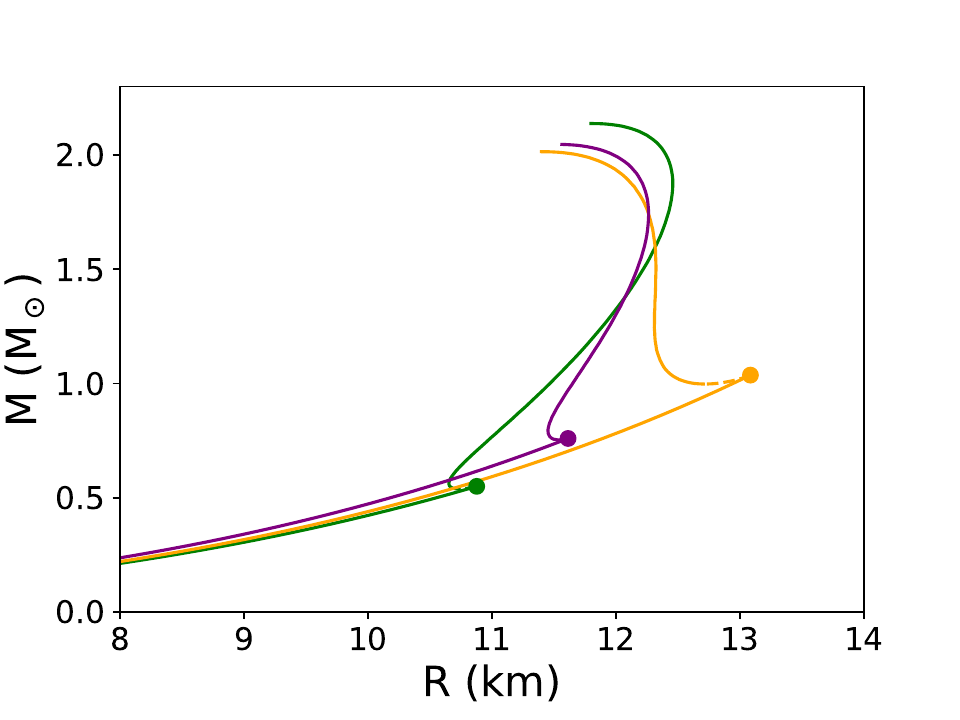}
    \caption{}
    \end{subfigure}
    \begin{subfigure}[b]{0.55\linewidth}
    \includegraphics[width=\linewidth]{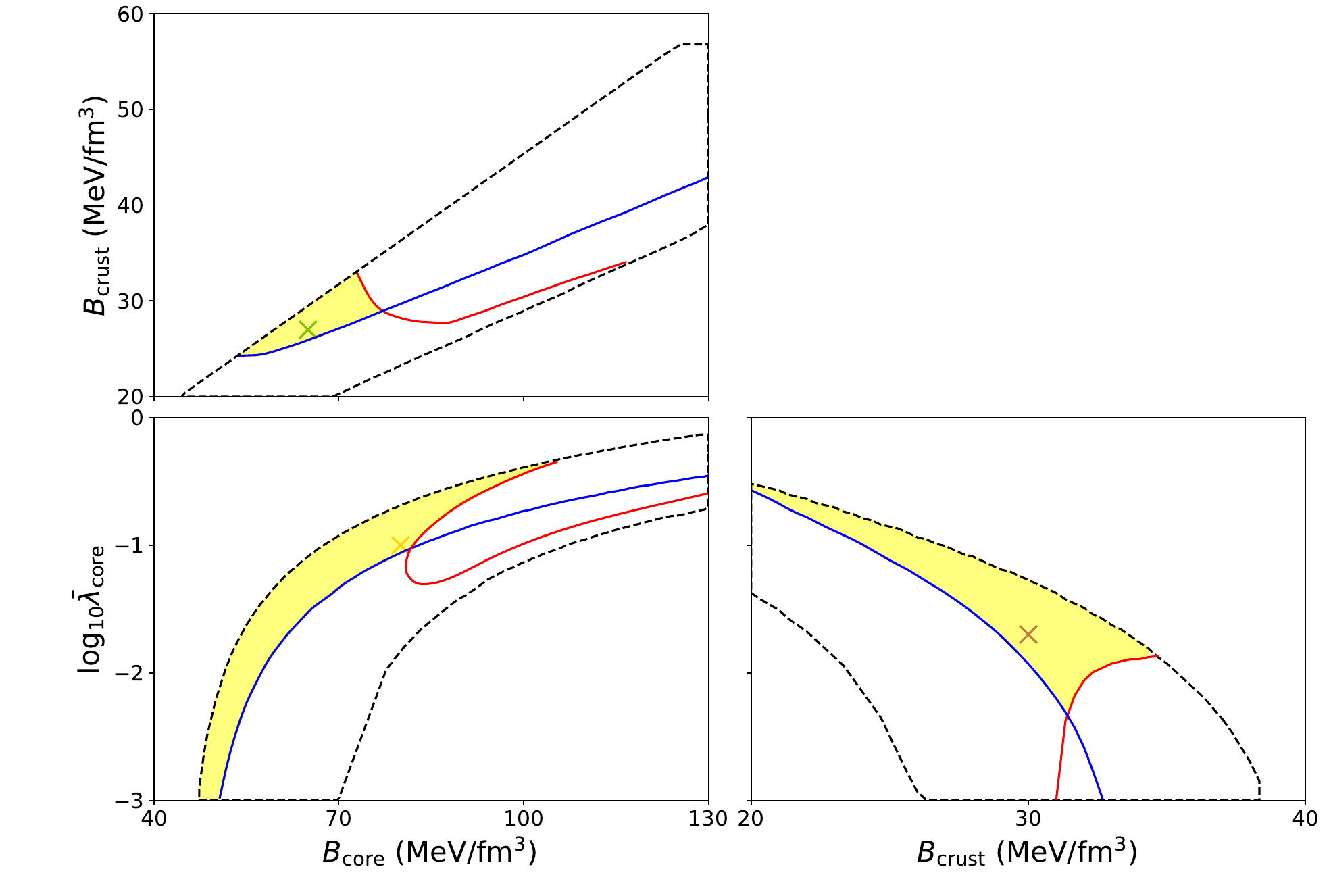}
    \caption{}
    \end{subfigure}
    \caption{(a) $M$-$R$ relations of HybQS benchmark examples with unpaired 2$f$ to 2SC+s or to CFL phase transition (these two cases have the same results). (b) The related viable parameter space (yellow-shaded region), where the cross markers correspond to the same-color $M$-$R$ in (a). In (b), the line-color convention follows that of Fig.~\ref{udQMtoSQM}b. Each panel is obtained by fixing one of the parameters, with the other two serving as the horizontal and vertical axes. For the top left panel, we choose $\bar{\lambda}_{\rm core}=0.06$, while for the two bottom panels, we choose $B_{\rm crust}=28\rm\, MeV/fm^{3}$ and $B_{\rm core}=65\rm \, MeV/fm^{3}$, respectively.}
    \label{udQMtoCFL}
\end{figure*}

Then we consider HybQSs with the unpaired 2$f$ to 2SC+s and unpaired 2$f$ to CFL phase transition cases. As aforementioned, for unpaired 2$f$ we have $\lambda_{\rm crust} = \bar{\lambda}_{\rm crust} = 0$, while Eq.~(\ref{types}) shows that the 2SC+s phase and CFL phase share the same $\xi_4$, and thus the same $P(\rho)$ and $\mu(P)$. Therefore, these two cases of phase transitions have degenerate results. The obtained mass-radius relations of benchmark examples are shown in Fig. \ref{udQMtoCFL}a, where we manage to identify twin star branch solutions, all of which are selected from a more extended viable parameter space illustration shown as the yellow shaded region in Fig.~\ref{udQMtoCFL}b, where the color superconductivity effects introduce an additional parameter $\bar{\lambda}_{\rm core}$. Besides, the left and right boundaries (black-dashed) in the $\bar{\lambda}_{\rm core}$-$B_{\rm core}$ panel are from considerations of relative stability at zero pressure and absence of stable HybQS branch, respectively, while the opposite order is in the $\bar{\lambda}_{\rm core}$-$B_{\rm crust}$ panel. In general, introducing $\bar{\lambda}_{\rm core}$ enlarges the viable parameter space, as most clearly shown by comparing Fig.~\ref{udQMtoCFL}b and Fig.~\ref{udQMtoSQM}b.

\subsection{2SC $\rightarrow$ 2SC$+$s and 2SC $\rightarrow$ CFL}
\begin{figure*}[htb!]
    \centering
    \includegraphics[width=.85\linewidth]{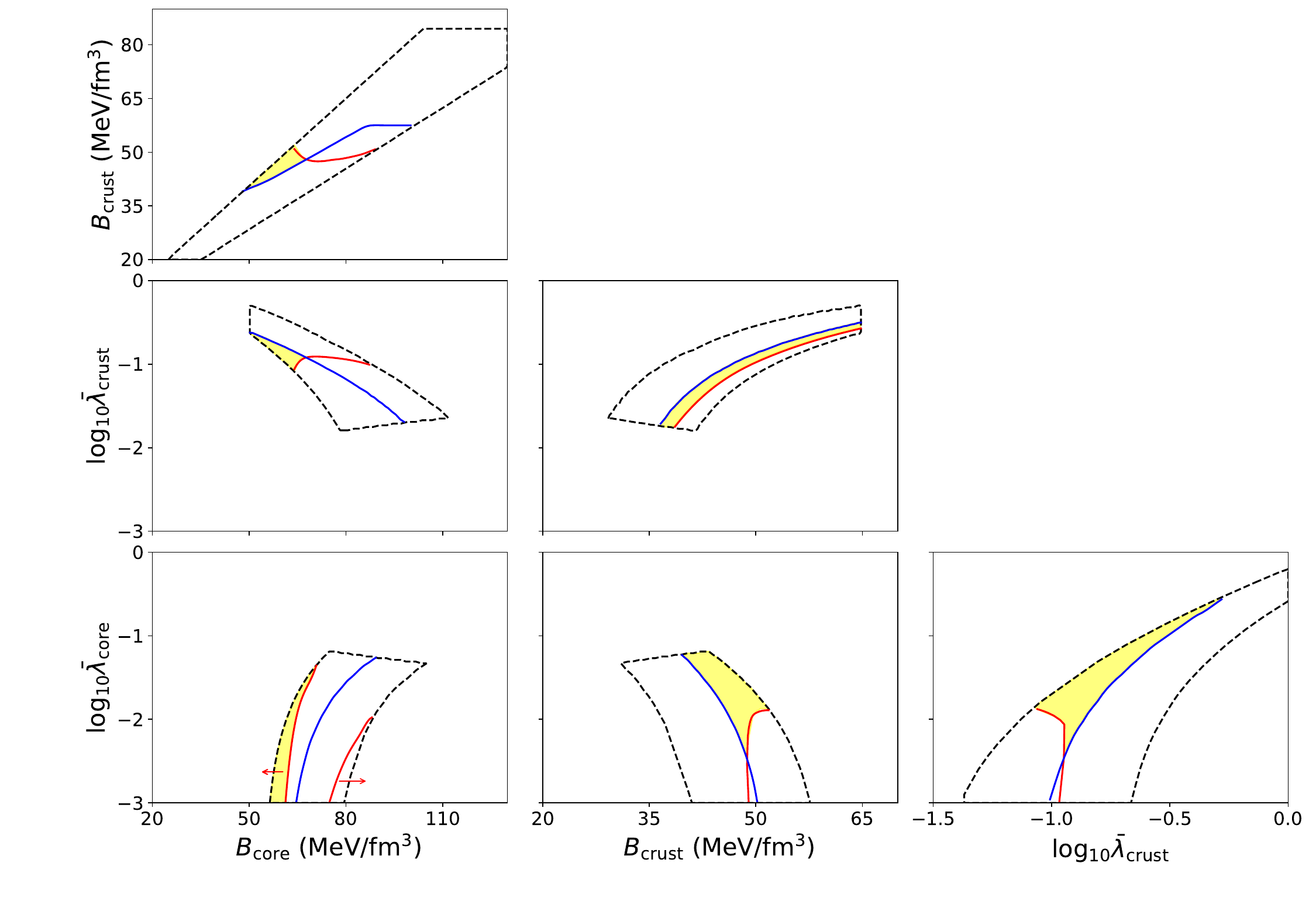}
 \includegraphics[width=.85\linewidth]{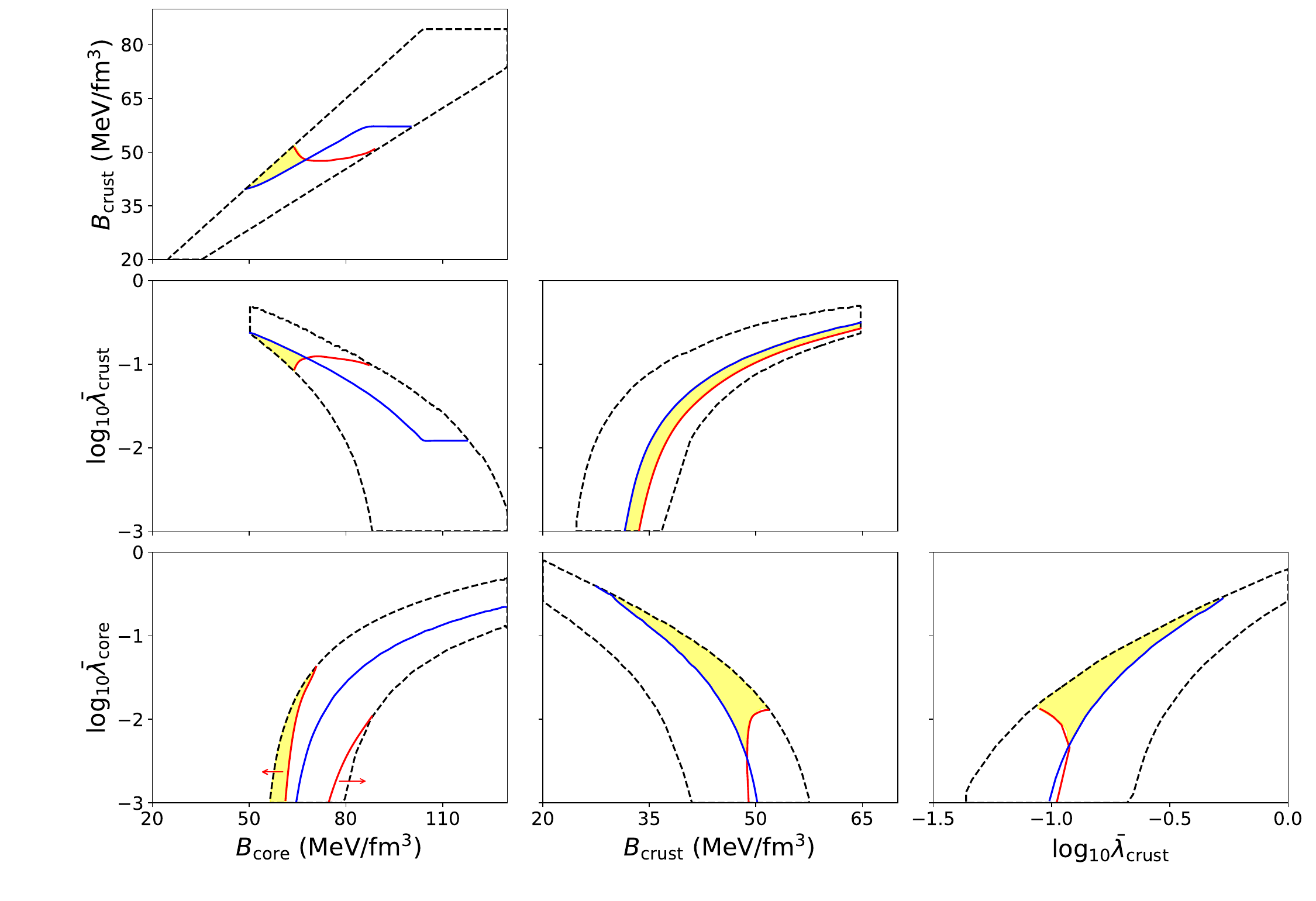}
    \caption{Parameter space of HybQSs with the 2SC to 2SC+s (top three rows) and the 2SC to CFL (bottom three rows) transitions. Each panel is obtained by fixing two of the parameters of ($B_{\rm crust}$, $B_{\rm core}$, $\bar{\lambda}_{\rm crust}$, $\bar{\lambda}_{\rm core}$)=(50 MeV/fm$^{3}$, 65 MeV/fm$^{3}$, 0.1, 0.01), with the other two serving as the horizontal and vertical axes correspondingly. The line-color convention follows that of Fig.~\ref{udQMtoSQM}b, with the additional red arrows in the bottom-left panels helping clarify the $M_{\rm TOV} \gtrsim 2M_{\odot}$ directions.}
    \label{combined}
\end{figure*}

We discuss the cases for HybQSs with the 2SC to CFL transition and the 2SC to 2SC+s transition in this one subsection since the results are quite similar, as can be seen in Fig.~\ref{combined}. The similarities between these two cases come from the fact that they share the same $P(\rho)$ and $\mu(P)$ given the same $\xi_4$, except that for the 2SC to 2SC+s case we have $\lambda_{\rm crust} > \lambda_{\rm core}$ always held, which sets the additional bottom and top black-dashed boundaries in $\bar{\lambda}_{\rm crust}$-$B$ and $\bar{\lambda}_{\rm core}$-$B$ related panels in the second and third rows of Fig.~\ref{combined}, respectively. In these two cases, early and late phase transitions give rise to two $M_{\rm TOV}=2M_{\odot}$ limit lines in the $\bar{\lambda}_{\rm core}$ versus $B_{\rm core}$ panels. Other features and analyses are similar to the other two cases discussed above, with twin star branches also identified in some subsets of parameter space.

\section{The deviation of the approximate universal relation between $f_{\rm peak}$ and $\Lambda_{1.35M_{\odot}}$}

Since whether or not a strong interaction phase transition occurs in compact stars is a crucial question for understanding the QCD phase diagram, it becomes natural and essential to search for observational signatures that could confirm its existence. Ref. \cite{bauswein2019identifying} performed simulations on merger of two $1.35\,M_\odot$ neutron stars and demonstrated that, if a phase transition occurs in the stellar interior, the dominant postmerger GW frequency $f_{\text{peak}}$ will deviate from the approximate universal relation between $f_{\text{peak}}$ and $\Lambda_{1.35M_{\odot}}$ (tidal deformability of $1.35M_\odot$ star). If the phase transition occurs above $1.35M_\odot$, $\Lambda_{1.35}$ remains unaffected. However, the softening of the EOS reduces the stellar radius and makes the remnant more compact, leading to a higher $f_{\text{peak}}$. As a result, $f_{\text{peak}}$ deviates from the approximate universal relation between $f_{\text{peak}}$ and $\Lambda_{1.35 M_{\odot}}$. While that study focused on the hadron-quark phase transitions, in this work, we investigate whether quark-quark phase transitions can also lead to similar deviations.

Since $f_{\text{peak}}$ is believed to be related to the fundamental oscillation mode ($f$-mode) frequency of the merger remnant \cite{rezzolla2016gravitational, shibata2005constraining, bauswein2012measuring, shibata2005merger, hotokezaka2013remnant}, there may exist an approximate universal relation between $f_{\text{peak}}$ and the $f$-mode frequency of neutron or quark stars. We choose the $f$-mode frequency ($f_{2M_{\odot}}$) of $2M_\odot$ star. In this case, a large quark core already appears at this mass and the phase transition affects $f_{\text{peak}}$ and $f_{2M_\odot}$ in a similar way, so the approximate universal relation between $f_{\text{peak}}$ and $f_{2M_\odot}$ may hold better compared with that between $f_{\text{peak}}$ and the $f$-mode frequency of lower-mass stars. We establish an approximate universal relation between $f_{\text{peak}}$ and $f_{2M_{\odot}}$, and then examine whether the phase transition leads to a deviation from the approximate universal relation between $f_{2M_{\odot}}$ and $\Lambda_{1.35M_{\odot}}$. In this way, possible deviations from the approximate universal relation between $f_{\text{peak}}$ and $\Lambda_{1.35M_{\odot}}$ can be inferred indirectly. 
\begin{figure}[h]
    \centering
    \includegraphics[width=1\linewidth]{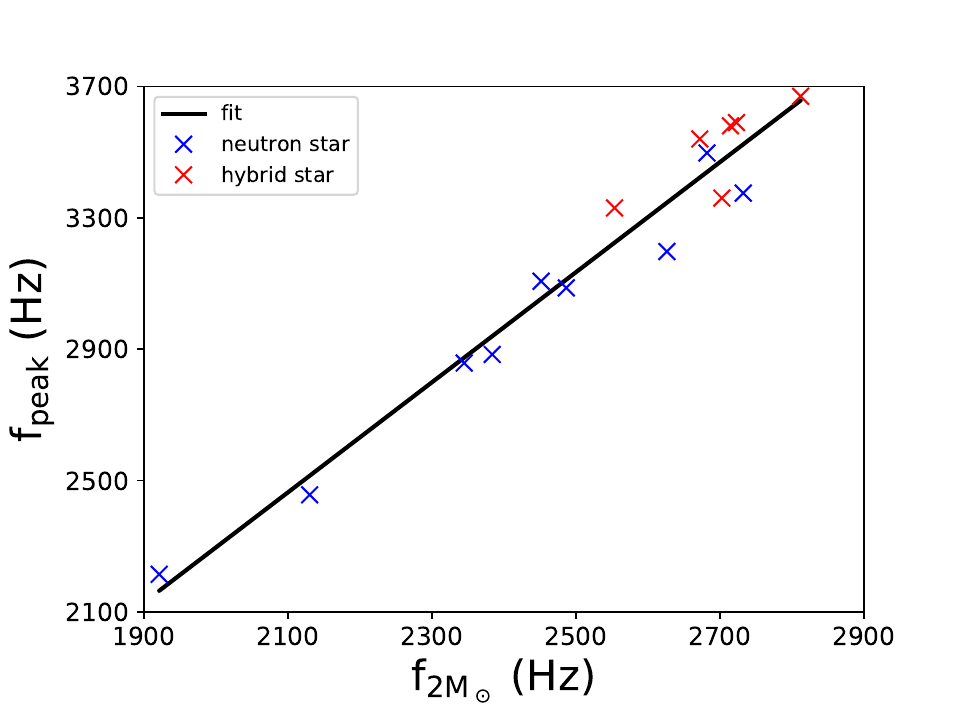}
    \caption{Approximate universal relation between the dominant postmerger frequency $f_{\text{peak}}$ and $f$-mode frequency of $2M_\odot$ star $f_{2M_{\odot}}$. Blue crosses indicate pure hadronic stars, red crosses indicate hadron–quark hybrid stars, and the black solid line shows the fitted relation between $f_{\text{peak}}$ and $f_{2M_{\odot}}$. }
    \label{fpeak_fmode}
\end{figure}

We employ both pure hadronic neutron star EOSs and hadron–quark hybrid star EOSs to investigate the relation between $f_{\text{peak}}$ and $f_{2M_{\odot}}$. The hadronic EOSs set includes model APR \cite{douchin2001unified}, BL \cite{bombaci2018equation}, DDHdelta \cite{grill2014equation}, SKa \cite{gulminelli2015unified}, SKb \cite{gulminelli2015unified}, SLY4 \cite{gulminelli2015unified}, SLY9 \cite{gulminelli2015unified} , GM1 \cite{xia2022unified}, NL3 \cite{xia2022unified1}, while the hybrid EOSs set includes different parameters of DD2F-SF model \cite{fischer2018quark}. All these EOSs are taken from Ref. \cite{Compose}. To calculate $f_{\text{peak}}$ of hadronic stars, we utilize Eq.~(\ref{fit}) by calculating $\Lambda_{1.35M_{\odot}}$ and substituting it into the relation \cite{bauswein2019identifying}:
\begin{equation}
\frac{f_\mathrm{peak}}{\rm kHz}=(6.486\times 10^{-7}\, \Lambda_{1.35M_{\odot}}^2 \,- \,2.231\times10^{-3}\,\Lambda_{1.35M_{\odot}}\,+\,4.1 ) .
\label{fit}
\end{equation}
For the $f$-mode frequency $f_{2M_{\odot}}$, we use the Cowling approximation and take the additional junction conditions at the discontinuous phase transition surface into consideration. In Fig.~\ref{fpeak_fmode}, we obtained an approximate universal relation between $f_{\text{peak}}$ and $f_{2M_{\odot}}$, which can be fitted by a linear function:
\begin{equation}
f_\mathrm{peak} = 1.675 \ f_{2M_{\odot}} - 1054\ \rm Hz\ .
\label{fit2}
\end{equation}
The approximate universal relation between $f_{\text{peak}}$ and $f_{2M_{\odot}}$ holds regardless of the presence of a first-order phase transition. Therefore, we infer that quark stars, whether undergoing a first-order phase transition or not, should also satisfy a similar relation.  
\begin{figure}[h]
    \centering
    \includegraphics[width=1\linewidth]{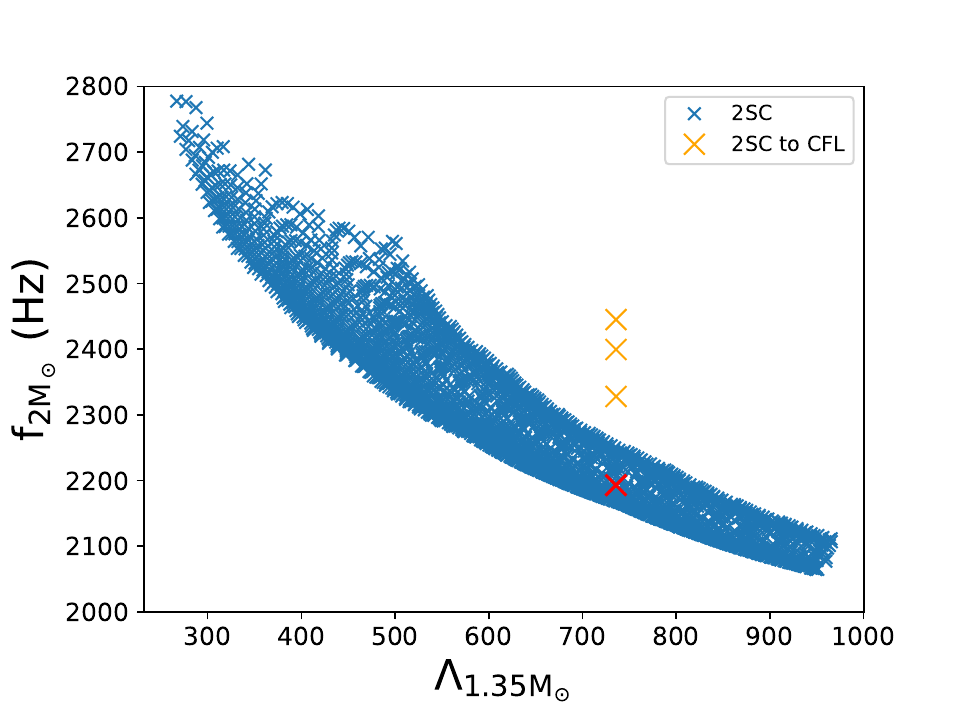}
    \caption{Relation between $f$-mode frequency of $2\,M_\odot$ star $f_{2M_{\odot}}$ and the tidal deformability of $1.35\,M_\odot$ star $\Lambda_{1.35M_{\odot}}$. Blue crosses represent pure 2SC quark stars of different parameters, and orange crosses correspond to 2SC to CFL HybQSs. Both pure 2SC quark stars and 2SC to CFL HybQSs shown in the figure satisfy $M_{\text{TOV}} \gtrsim 2M_\odot$ and $\Lambda_{1.4M_{\odot}} < 800$. Red crosses denote the pure 2SC quark star corresponding to the orange crosses, but without the phase transition. The top orange cross shows a deviation of about 220Hz from the approximate universal relation. The parameters of the cross showing the maximum deviation are ($B_{\rm crust}$, $B_{\rm core}$, $\bar{\lambda}_{\rm crust}$, $\bar{\lambda}_{\rm core}$)=(83 MeV/fm$^{3}$, 93 MeV/fm$^{3}$, 0.756, 0.195). The parameters of the red crosses are ($B$, $\bar{\lambda}$)=(83 MeV/fm$^{3}$, 0.756).}
    \label{2SCtoCFLdeviation}
\end{figure}
Then, we calculate $f_{2M_{\odot}}$ and $\Lambda_{1.35M_{\odot}}$ for both the pure 2SC phase quark stars and the 2SC to CFL HybQSs, as shown in Fig.~\ref{2SCtoCFLdeviation}. We find the presence of phase transitions leads to a deviation of $f_{2M_{\odot}}$ from the approximate universal relation between $f_{2M_{\odot}}$ and $\Lambda_{1.35M_{\odot}}$. Combining the approximate universal relation between $f_{\text{peak}}$ and $f_{2M_\odot}$, we indirectly suggest that quark–quark phase transitions may also cause $f_{\text{peak}}$ to deviate from the approximate universal relation between $f_{\text{peak}}$ and $\Lambda_{1.35M_\odot}$. Since the maximum deviation of $f_{2M_{\odot}}$ can reach 220 Hz, combining with Eq. (\ref{fit2}), we obtain that the deviation of $f_{\text{peak}}$ is about 368 Hz. The deviation is similar to that found in hadron–quark hybrid stars \cite{bauswein2019identifying}, which could be detected in future observations and used to identify whether a first-order phase transition exists in compact stars.

\section{SUMMARY and outlook}
With the simple MIT+CSS construction, we found that the generic $M$-$R$ phase diagram for the hybrid quark star branches shares similar features with conventional hybrid stars. Then, with a more phenomenological quark matter model that encompasses various phases of quark matter, we explored the new possibility of quark-quark phase transitions within quark stars and demonstrated the viability of the resulting HybQSs under conventional astrophysical constraints. Specifically, we constructed models for HybQSs and obtained their EOSs associated with different quark-quark transition types by examining different quark matter phases and their relative stabilities under varying EOS parameters. Models for transitions between phases with the same quark flavors were found to be non-viable from our modelling perspective. Applying astrophysical constraints from the pulsar mass measurement ($M_{\mathrm{TOV}} \gtrsim 2 M_{\odot}$) and the tidal deformability ($\Lambda_{1.4M_{\odot}} < 800$) from GW170817, we identified viable parameter regions for the remaining five models where phase transitions are accompanied by flavor-composition change. We also found that the introduction of color superconductivity can enlarge the viable parameter space and enable twin-star solutions that satisfy the astrophysical constraints. We find that an approximate universal relation exists between $f_{\text{peak}}$ and $f_{2M_{\odot}}$ regardless of the presence of phase transitions. Meanwhile, we find that quark–quark phase transitions cause $f_{2M_{\odot}}$ to deviate from the approximate universal relation between $f_{2M_{\odot}}$ and $\Lambda_{1.35M_{\odot}}$. Combining these results, we demonstrated that quark-quark phase transitions may also lead $f_{\text{peak}}$ to deviate from the approximate universal relation between $f_{\text{peak}}$ and $\Lambda_{1.35M_{\odot}}$.

This new type of hybrid stellar object can have rich astrophysical signatures. The GWs related to its nonradial oscillations~\cite{Zhang:2023zth,Sun:2025zpj,Li:2024hzt,Guha:2024gfe}, the approximate universal relations~\cite{Yagi:2013bca,Largani:2021hjo,Zhao:2025pgx,Pretel:2024pem}, 
the post-merger dynamics~\cite{Weih:2019xvw,Shibata:2019ctb,Zhou:2024syq}, and the electromagnetic signatures~\cite{Menezes:2006zx,Geng:2021apl,wang2018frb,Wang:2024opz,Xu:2025wwb}, all may manifest differently from those of neutron stars and conventional hybrid stars. Besides, it is also interesting to examine this new type of objects in explaining other unconventional observations, such as HESS J1731-347 and XTE J1814-338~\cite{Zhang:2025rnf}. 

In this proof-of-concept study, we made several simplifications for convenience, considering the large uncertainties of non-perturbative QCD dynamics. For example, we ignored the perturbative QCD corrections and the density dependence of the bag constant, all of which may alter the types of quark-quark phase transitions and associated viable parameter space. Besides, it is also interesting to explore the possibilities of mixed phases (Gibbs construction)~\cite{schertler1999neutron,Constantinou:2023ged}, crossover~\cite{fukushima2010phase,masuda2013hadron,Blaschke:2021poc,Yuan:2023dco,Li:2018ayl}, sequential multi-phase transitions~\cite{Alford:2017qgh,Li:2023zty}, and phase-transition-induced collapse~\cite{lin2006gravitational,yip2023gravitational,huang2024explosion}. A general Bayesian analysis of the whole parameter space is also worth exploring. Moreover, the dominant postmerger frequency $f_{\text{peak}}$ could be determined more accurately through numerical simulations of binary quark star mergers. We leave these more exhaustive explorations for future studies.

\acknowledgements{We thank Yurui Zhou, Duanyuan Gao, and Jing Ren for useful discussions. This work is supported by the National SKA Program of China No. 2020SKA0120300 and NSFC Grant No. 12203017. C.~Zhang is supported by the Fundamental Research Funds for the
Central Universities and the Jockey Club Institute for Advanced Study at The Hong Kong University of Science and Technology. W.-L. Yuan is supported by the Special Funds of the National Natural Science Foundation of China (Grant No. 12447171) and the China Postdoctoral Science Foundation (Grant No. 2025M773418).}

\bibliography{references1.bib}

\end{document}